\documentclass[fleqn,usenatbib]{mnras}

\usepackage{newtxtext,newtxmath}

\usepackage[T1]{fontenc}

\DeclareRobustCommand{\VAN}[3]{#2}
\let\VANthebibliography\thebibliography
\def\thebibliography{\DeclareRobustCommand{\VAN}[3]{##3}\VANthebibliography}

\usepackage{graphicx}	
\usepackage{amsmath}	

\title[Swift J170800$-$402551.8]{Swift Deep Galactic Plane Survey Classification of Swift J170800$-$402551.8 as a Candidate Intermediate Polar Cataclysmic Variable}

\author[B. O'Connor et al.]{
B. O'Connor,$^{1,2,3}$\thanks{E-mail: oconnorb@gwmail.gwu.edu}
E. G\"{o}\u{g}\"{u}\c{s},$^{4}$ 
J. Hare,$^{3,5,6}$ 
K. Mukai,$^{7,8}$ 
D. Huppenkothen,$^9$ 
J. Brink,$^{10,11}$ 
\newauthor 
D. A. H. Buckley,$^{11,12,10,13}$ 
A. Levan,$^{14}$
M. G. Baring,$^{15}$
R. Stewart,$^1$ 
C. Kouveliotou,$^1$
P. Woudt,$^{10}$ 
\newauthor 
E. Bellm,$^{16}$ 
S. B. Cenko,$^{3,17}$
P. A. Evans,$^{18}$
J. Granot,$^{19,20,1}$ 
C. Hailey,$^{21}$
F. Harrison,$^{22}$ 
D. Hartmann,$^{23}$ 
\newauthor
A. J. van der Horst,$^1$
L. Kaper,$^{24}$
J. A. Kennea,$^{25}$
S. B. Potter,$^{11,26}$
P. O. Slane,$^{27}$
D. Stern,$^{28}$
\newauthor
R. A. M. J. Wijers,$^{29,1}$
G. Younes$^{3,1}$
\\
$^{1}$Department of Physics, The George Washington University, Washington, DC 20052, USA\\
$^{2}$Department of Astronomy, University of Maryland, College Park, MD 20742-4111, USA\\
$^{3}$Astrophysics Science Division, NASA Goddard Space Flight Center, 8800 Greenbelt Rd, Greenbelt, MD 20771, USA\\
$^{4}$Sabanc\i~University, Faculty of Engineering and Natural Sciences, \.Istanbul 34956 Turkey\\
$^{5}$Center for Research and Exploration in Space Science and Technology, NASA/GSFC, Greenbelt, Maryland 20771, USA \\
$^{6}$The Catholic University of America, 620 Michigan Ave., N.E. Washington, DC 20064, USA \\
$^{7}$CRESST II and X-ray Astrophysics Laboratory, NASA/GSFC, Greenbelt, MD 20771, USA \\
$^{8}$Department of Physics, University of Maryland Baltimore County, 1000 Hilltop Circle, Baltimore MD 21250, USA \\
$^{9}$SRON Netherlands Institute for Space Research, Niels Bohrweg 4, 2333CA Leiden, The Netherlands \\
$^{10}$Department of Astronomy, University of Cape Town, Private Bag X3, Rondebosch 7701, South Africa \\
$^{11}$South African Astronomical Observatory, P.O. Box 9, Observatory 7935, Cape Town, South Africa \\
$^{12}$Southern African Large Telescope, P.O. Box 9, Observatory 7935, Cape Town, South Africa \\ 
$^{13}$Department of Physics, University of the Free State, P.O. Box 339, Bloemfonein 9300, South Africa \\
$^{14}$Department of Astrophysics/IMAPP, Radboud University, P.O. Box 9010, 6500 GL, The Netherlands \\
$^{15}$Department of Physics and Astronomy - MS 108, Rice University, 6100 Main Street, Houston, Texas 77251-1892, USA \\
$^{16}$DIRAC Institute, Department of Astronomy, University of Washington, 3910 15th Avenue NE, Seattle, WA 98195, USA \\
$^{17}$Joint Space-Science Institute, University of Maryland, College Park, MD 20742 USA \\
$^{18}$School of Physics and Astronomy, University of Leicester, University Road, Leicester, LE1 7RH, UK \\
$^{19}$Department of Natural Sciences, The Open University of Israel, P.O Box 808, Ra'anana 43537, Israel \\
$^{20}$Astrophysics Research Center of the Open university (ARCO), The Open University of Israel, P.O Box 808, Ra’anana 43537, Israel \\
$^{21}$Columbia Astrophysics Laboratory, Columbia University, New York, NY 10027, USA \\
$^{22}$Cahill Center for Astrophysics, California Institute of Technology, 1216 East California Boulevard, Pasadena, CA 91125, USA \\
$^{23}$Department of Physics and Astronomy, Clemson University, Kinard Lab of Physics, Clemson, SC 29634-0978, USA \\
$^{24}$University of Amsterdam, Science Park 904, 1098 XH Amsterdam, The Netherlands \\
$^{25}$Department of Astronomy and Astrophysics, The Pennsylvania State University, 525 Davey Lab, University Park, PA 16802, USA \\
$^{26}$Department of Physics, University of Johannesburg, PO Box 524, Auckland Park 2006, South Africa \\
$^{27}$Center for Astrophysics, Harvard \& Smithsonian, 60 Garden St. Cambridge, MA 02138, USA \\
$^{28}$Jet Propulsion Laboratory, California Institute of Technology, 4800 Oak Grove Drive, Mail Stop 169-221, Pasadena, CA 91109, USA \\
$^{29}$Anton Pannekoek Institute, University of Amsterdam, Postbus 94249, 1090 GE Amsterdam, The Netherlands \\
}

\date{Accepted XXX. Received YYY; in original form ZZZ}

\pubyear{2023}

\begin{document}
\label{firstpage}
\pagerange{\pageref{firstpage}--\pageref{lastpage}}
\maketitle

\begin{abstract}
Here, we present the results of our multi-wavelength campaign aimed at classifying \textit{Swift} J170800$-$402551.8 as part of the \textit{Swift} Deep Galactic Plane Survey (DGPS). We utilized Target of Opportunity (ToO) observations with \textit{Swift}, \textit{Chandra}, \textit{NICER}, 
\textit{XMM-Newton}, \textit{NuSTAR}, and the Southern African Large Telescope (SALT), as well as multi-wavelength archival observations from \textit{Gaia}, VPHAS, and VVV. The source displays a periodicity of 784 s in our \textit{XMM-Newton} observation. The X-ray spectrum (\textit{XMM-Newton} and \textit{NuSTAR}) can be described by thermal bremsstrahlung radiation with a temperature of $kT$\,$\approx$\,$30$ keV. The phase-folded X-ray lightcurve displays a double-peaked, energy-dependent pulse-profile. We used \textit{Chandra} to precisely localize the source, allowing us to identify and study the multi-wavelength counterpart. Spectroscopy with SALT identified a Balmer H$\alpha$ line, and potential HeI lines, from the optical counterpart. The faintness of the counterpart ($r$\,$\approx$\,$21$ AB mag) favors a low-mass donor star.  Based on these criteria, we classify \textit{Swift} J170800$-$402551.8 as a candidate intermediate polar cataclysmic variable, where the spin period of the white dwarf is 784 s. 
\end{abstract}




\section{Introduction}
\label{sec: intro}

The Galactic X-ray sky harbors a variety of high-energy sources including magnetars, low-mass X-ray binaries (LMXBs), high-mass X-ray binaries (HMXBs), neutron stars (NS), and cataclysmic variables (CVs). Each of these source classes display distinct behaviors, but overlaps in a few common properties, such as their hard X-ray emission, periodicity, and optical emission line spectra. The confident classification of these sources is challenging for the faint X-ray population (i.e., X-ray fluxes below $10^{-11}$ erg cm$^{2}$ s$^{-1}$), and can, in many cases, rely on serendipitously detecting outbursts or flaring activity. Many of these sources are classified based on their X-ray behavior (e.g., magnetars, LMXBs, HMXBs; \citealt{Kaspi2017,Kretschmar2019}), while others are largely found in optical surveys (e.g. CVs; \citealt{Oliveira2020,Szkody2020}). 

The \textit{Swift} Deep Galactic Plane Survey (DGPS; \citealt{DGPS}) utilized the \textit{Neil Gehrels Swift Observatory} \citep{Gehrels2004} X-ray Telescope \citep[XRT;][]{Burrows2005} to conduct a survey of 40 deg$^2$ of the Galactic Plane (GP). The Survey covered regions of Galactic longitude $10$\,$<$\,$|l|$\,$<$\,$30$ deg and latitude $|b|$\,$<$\,$0.5$ deg. In total, the Survey is made up of 380 XRT tiles, each having a duration of 5 ks. The Survey identified 928 unique X-ray sources, of which 348 ($\sim$\,40\%) were previously unknown. 
We have carried out multiple extensive follow-up campaigns of unclassified sources found in the Survey. These include a LMXB \citep{Gorgone2019}, an intermediate polar (IP) CV \citep{Gorgone2021}, a polar CV \citep{OConnor2023polar}, and a Be X-ray binary (BeXRB; \citealt{OConnor2021}). 

CVs are interacting binaries comprising a white dwarf (WD) in an orbit with a main sequence star \citep{Warner1995}. If the magnetic field of the WD is strong ($>$\,1 MG) then the systems are referred to as magnetic CVs, and are generally strong X-ray emitters \citep{deMartino2020,Shaw2020}. Magnetic CVs come in two flavors: polars (e.g., AM Herculis; \citealt{Cropper1990}) or intermediate polars (e.g., DQ Herculis; \citealt{Patterson1994}). In polar CVs the magnetic field strength is strong enough ($>$\,10 MG) to fully disrupt the formation of an accretion disk \citep{Ramsay2007}, and also causes the WD spin and orbital periods to become locked (synchronous rotation). Whereas in IP systems, the accretion disk is only partially disrupted \citep{Mukai2017}, and the weaker magnetic field does not lock the motion of the system (asynchronous rotation; \citealt{Norton2004}). In these systems, the spin period of the WD is instead usually significantly shorter than the orbital period of the binary. IPs further separate themselves from polars through their more luminous, and less variable, X-ray emission \citep{Mukai2017}.

Here, we present our multi-wavelength analysis (X-ray, optical, and infrared) 
of \textit{Swift} J170800$-$402551.8, hereafter referred to as J1708. 
The source was first detected with \textit{Swift} in 2012 \citep{Reynolds2013}, despite earlier (less sensitive) X-ray observations from \textit{ASCA} and \textit{ROSAT} covering its position. We carried out ToO observations with \textit{XMM-Newton}, \textit{NuSTAR}, and \textit{NICER} to study the X-ray spectrum and search for coherent pulsations. We also utilized a \textit{Chandra} ToO to localize the optical and infrared counterpart, which we then observed with SALT. We further analyzed infrared lightcurves of the counterpart to search for variability, and compiled available optical and infrared photometry to study the spectral energy distribution (SED). 

Based on these data, we determine that J1708 is likely an accreting binary and we consider multiple interpretations for the source as either an HMXB, LMXB, or magnetic CV. 
We further discuss our observations and data analysis in \S \ref{sec: obs/analysis}. The results are discussed in \S \ref{sec: results}, and our interpretation and conclusions regarding the source classification are presented in \S \ref{sec: discussion} and \S \ref{sec: conclusions}. Error bars are reported at the $1\sigma$ level and upper limits at the $3\sigma$ level, unless otherwise stated.

\section{Observations and Data Analysis}
\label{sec: obs/analysis}

\subsection{X-ray Data}

\begin{table*}
\centering
\caption{Log of X-ray observations used in this work, including from \textit{Swift}, \textit{NICER}, \textit{Chandra}, \textit{NuSTAR}, and \textit{XMM-Newton}. Observations after 2012 were obtained through DGPS operations \citep{DGPS} or through approved DGPS follow-up programs.}
\label{tab: observations}
\begin{tabular}{lccccc}
\hline\hline
\textbf{Start Time (UT)} & \textbf{Telescope} & \textbf{Instrument} &  \textbf{Exposure (s)} & \textbf{ObsID}  \\
\hline
 2012-07-22 09:58:59 & \textit{Swift} & XRT  & 613 &
 00043282001 \\
 2012-07-26 21:24:59 & \textit{Swift} & XRT  & 634 &
 00043289001\\
 2020-01-26 02:59:37 & \textit{Swift} & XRT  &4693  &03110768001
 \\
  2020-03-24 05:01:09 & \textit{NuSTAR} & FPMA/B & 41334 & 90601310002\\
 2020-03-24 06:43:47 & \textit{XMM-Newton} & MOS/PN &  40800 & 0821860301 \\
 2020-03-24 11:33:35 & \textit{Swift} & XRT  &  1650 & 00089032001
 \\
  2020-05-21 21:57:36& \textit{Swift} & XRT  & 82 & 03110768002
 \\
 2020-06-09 21:49:35 & \textit{Swift} & XRT  & 245 & 03110768003
 \\
 2022-03-07 12:20:06  & \textit{Chandra} & ACIS-I  & 2580  & 25174
 \\
2022-03-07 20:11:36 & \textit{NICER} & XTI  & 5298 & 4550030101 \\
  2022-03-08 05:29:34  & \textit{NICER} & XTI  & 9488 & 4550030102 \\
\hline\hline
\end{tabular}
\end{table*}

\subsubsection{\textit{Swift}}
\label{sec: XRT}

The field of J1708 has been observed by the \textit{Swift} X-ray Telescope (XRT) on six occasions since 2012 (Table \ref{tab: observations}), including as part of the DGPS \citep{DGPS}. We analyzed these data using the methods outlined by \citet{DGPS}. In these observations, J1708 displays a mean count rate of $(2.0\pm0.2)\times10^{-2}$ cts s$^{-1}$. No flaring or outbursts were discovered in the archival lightcurve. Utilizing these XRT observations, we derive an enhanced position \citep{Evans2009} of RA, DEC (J2000) = $17^{h}08^m 00^{s}.30$, $-40^\circ 25\arcmin 54.9\arcsec$ with an accuracy of $2.4\arcsec$ (90\% CL). 

\subsubsection{\textit{XMM-Newton}}
\label{sec: XMM}

We used one of our approved \textit{XMM-Newton} ToOs to observe J1708 (ID: 082186; PI: Kouveliotou) for 40 ks with the EPIC-pn and MOS (MOS1/MOS2) cameras in Full Frame mode using the medium filter. The data were reduced and analyzed using tasks with the Science Analysis System (\texttt{SAS v18.0.0}) software. 
We extracted the source photons in a circular region ($30\arcsec$) identified by the \texttt{eregionanalyse} task. 
The background photons were taken from an annulus surrounding the source with radii of $60\arcsec$ and $100\arcsec$, respectively. Spectra were grouped to a minimum of 1 count per bin. The response matrix and ancillary response files were obtained using the \texttt{rmfgen} and \texttt{arfgen} tasks, respectively. The event arrival times were barycenter corrected to the solar system using the \texttt{barycen} task. 

\subsubsection{\textit{NuSTAR}}
\label{sec:  nustar}

A 40 ks \textit{NuSTAR} observation of J1708 was carried out simultaneously to our \textit{XMM-Newton} ToO (see Table \ref{tab: observations}) through the DGPS \textit{NuSTAR} Legacy Survey Program\footnote{\url{https://www.nustar.caltech.edu/page/59\#g9}}. We used standard tasks within the \textit{NuSTAR} Data Analysis Software pipeline (\texttt{NuSTARDAS v2.1.1}) within \texttt{HEASoft v6.29c} to extract lightcurves and spectra from a circular region of size 60\arcsec{} from both FPMA and FPMB. The arrival time of photons were barycenter-corrected to the solar system\footnote{\url{https://heasarc.gsfc.nasa.gov/ftools/caldb/help/barycorr.html}}. The spectra were grouped to a minimum of 15 counts per bin.

\subsubsection{\textit{Chandra}}
\label{sec: chandra}

We used one of our approved \textit{Chandra} Cycle 23 ToOs (ID: 23500070; PI: Kouveliotou) to target J1708 (ObsID: 25174) for a total of 2.5 ks exposure. The \textit{Chandra} data were retrieved from the 
\textit{Chandra} Data Archive (CDA)\footnote{\url{https://cda.harvard.edu/chaser/}}. We re-processed the data using the \texttt{CIAO v4.14}  data reduction package with \texttt{CALDB} Version 4.9.6, and filtered the events to the energy range $0.5-7$ keV. Source detection was performed using the \texttt{wavdetect} task. J1708 is localized to RA, DEC (J2000) = $17^{h}08^m 00^{s}.43$, $-40^\circ 25\arcmin 53.86\arcsec$. Due to the short exposure, there are no other sources detected. 
Given the lack of other sources in the field, we cannot correct the native astrometry. We adopt a systematic uncertainty of $0.8\arcsec$ at the 90\% confidence level\footnote{\url{https://cxc.cfa.harvard.edu/cal/ASPECT/celmon/}} for the source position. 

\subsubsection{\textit{NICER}}
\label{sec: nicer}

\textit{NICER} observations of J1708 were obtained under our Cycle 3 program (ID: 4050; PI: Kouveliotou) using the X-ray timing instrument \citep[XTI;][]{Gendreau2016}. 
We analyzed two NICER observations (ObsIDs: 4550030101 and 4550030102) using \texttt{NICERDAS v.10a} and \texttt{CALDB xti20221001} from \texttt{HEASOFT v.6.31.1}. After retrieving the latest geomagnetic data with the task \texttt{geomag}, we processed the raw mission data using \texttt{nicerl2} to generate filtered, cleaned event files and barycentered them with \texttt{barycorr}. We then used \texttt{nicerl3-spect} to produce the spectral files and \texttt{SCORPEON} background models. The source is only marginally detected in these observations. We note that a prominent Oxygen K$\alpha$ emission line is visible in the spectra, which is due to Earth's atmosphere and not intrinsic to our source (K. Gendreau, private communication). 


\subsection{Optical/infrared Data}

 \begin{figure}
\centering
\includegraphics[width=\columnwidth]{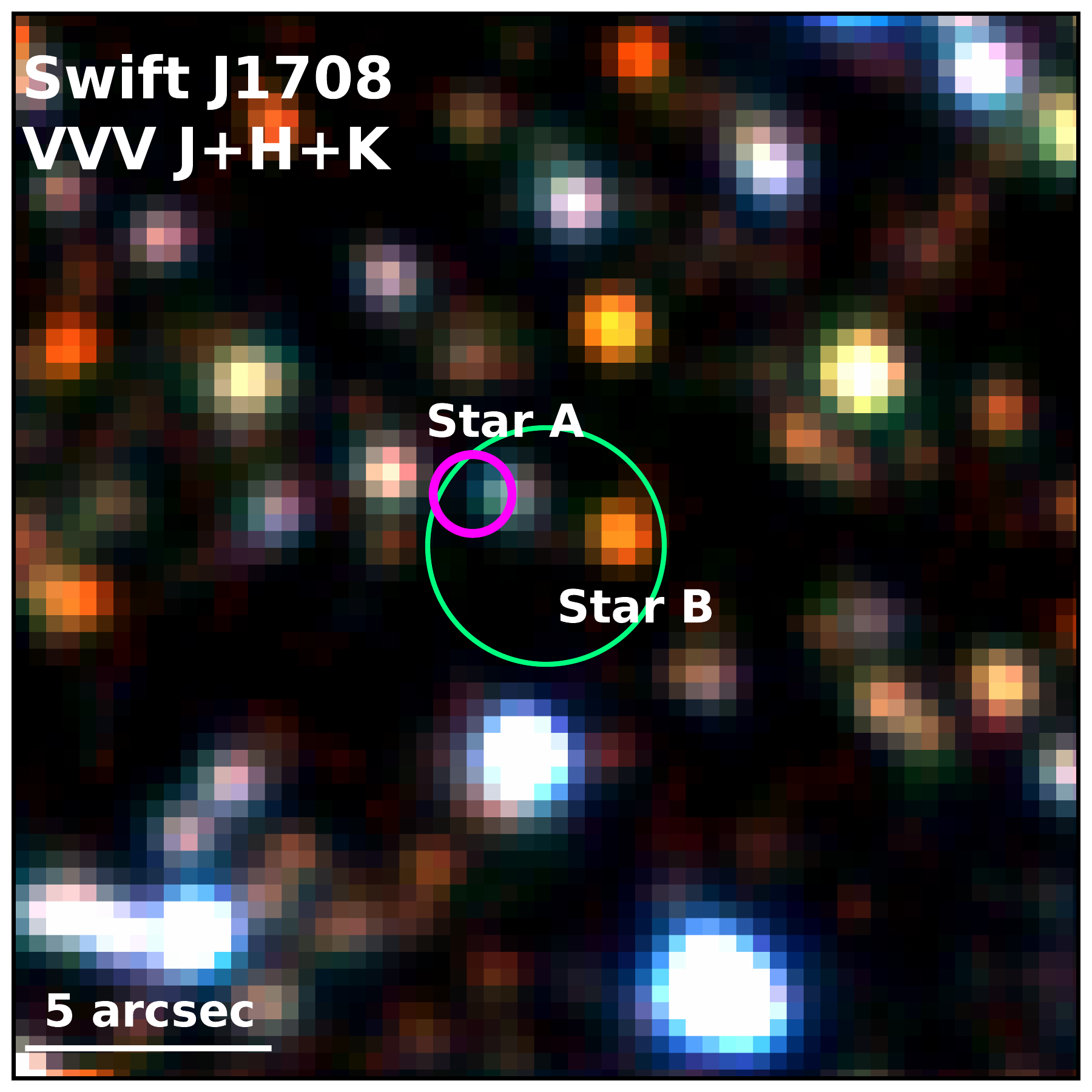}
\caption{RGB image of the field of J1708 using three VVV filters: red = $K_s$, green = $H$, and blue = $J$. The XRT enhanced position with radius $2.4\arcsec$ is displayed by a light green circle. The \textit{Chandra} localization ($0.8\arcsec$) is displayed by the magenta circle, and favors Star A as the counterpart. 
North is up and East is to the left. 
}
\label{fig: counterpart}
\end{figure}

\subsubsection{Archival Multi-wavelength Data}
\label{sec: archival}

We searched the \texttt{VizieR}\footnote{\url{http://vizier.cds.unistra.fr/viz-bin/VizieR}} Database \citep{2000A&AS..143...23O} for an optical and infrared source consistent with the X-ray localization of J1708. We identified two stars within the XRT localization. These sources are found in \textit{Gaia} EDR3 \citep{gaiaEDR3}, the VIRAC catalog \citep{Smith2018} of the VISTA Variables in the Via Lactea (VVV) survey, and the VST Photometric $H\alpha$ Survey \citep[VPHAS;][]{Drew2014,Drew2016} catalog: \textit{i}) Star A (\textit{Gaia} source ID: 5966886124607467136) is a blue source residing in the North-East portion of the XRT position and \textit{ii}) Star B is a very red source only detected in the infrared in the VVV Survey. We note that Star A is consistent with the \textit{Chandra} localization of J1708 (\S \ref{sec: chandra}), which disfavors an association to Star B. 

A finding chart using the VVV data is visible in Figure \ref{fig: counterpart} and both stars are labeled for clarity. We note that neither source is detected with \textit{Swift}/UVOT or the \textit{XMM-Newton} Optical/UV Monitor Telescope, likely due to dust in our Galaxy. We include the archival photometry (AB system) of Star A in Table \ref{tab: optcounterpart}. 

The VPHAS photometry indicates that the source is likely a strong $H\alpha$ emitter. Therefore, we performed optical spectroscopy (see \S \ref{sec: salt}) to confirm the presence of $H\alpha$.

\begin{table}
    \centering
    \caption{Photometry of the optical/infrared counterpart to J1708. The magnitudes $m_\lambda$ are reported in the AB magnitude system. Limiting magnitudes from VPHAS are at the $5\sigma$ level \citep{Drew2014}.
    }
    \label{tab: optcounterpart}
    \begin{tabular}{lcccc}
    \hline
    \hline
\textbf{Source} & \textbf{Filter}  & \textbf{$m_\lambda$} \textbf{(mag)}   \\
    \hline
  \textit{Gaia}  & $G$ & $20.67\pm0.02$ \\
  \textit{Gaia}  & $G_{BP}$ & $21.6\pm0.2$ \\
  \textit{Gaia}  & $G_{RP}$ & $19.73\pm0.10$ \\
  VPHAS  & $u$ & $>21.7$ \\
  VPHAS  & $g$  & $>22.8$ \\
  VPHAS & $H\alpha$ & $19.59\pm0.06$ \\
  VPHAS & $r$ & $20.93\pm0.10$ \\
  VPHAS & $i$ & $20.06\pm0.05$ \\
  VIRAC & $Z$ & $19.90\pm0.15$ \\
  VIRAC & $Y$ & $19.1\pm0.2$ \\
  VIRAC & $J$  & $18.92\pm0.15$ \\
  VIRAC & $H$ & $18.02\pm0.10$ \\
  VIRAC & $K$ & $18.14\pm0.15$\\
    \hline
    \hline
    \end{tabular}
\end{table}

\subsubsection{VVV Lightcurve}

\begin{figure}
\centering
\includegraphics[width=\columnwidth]{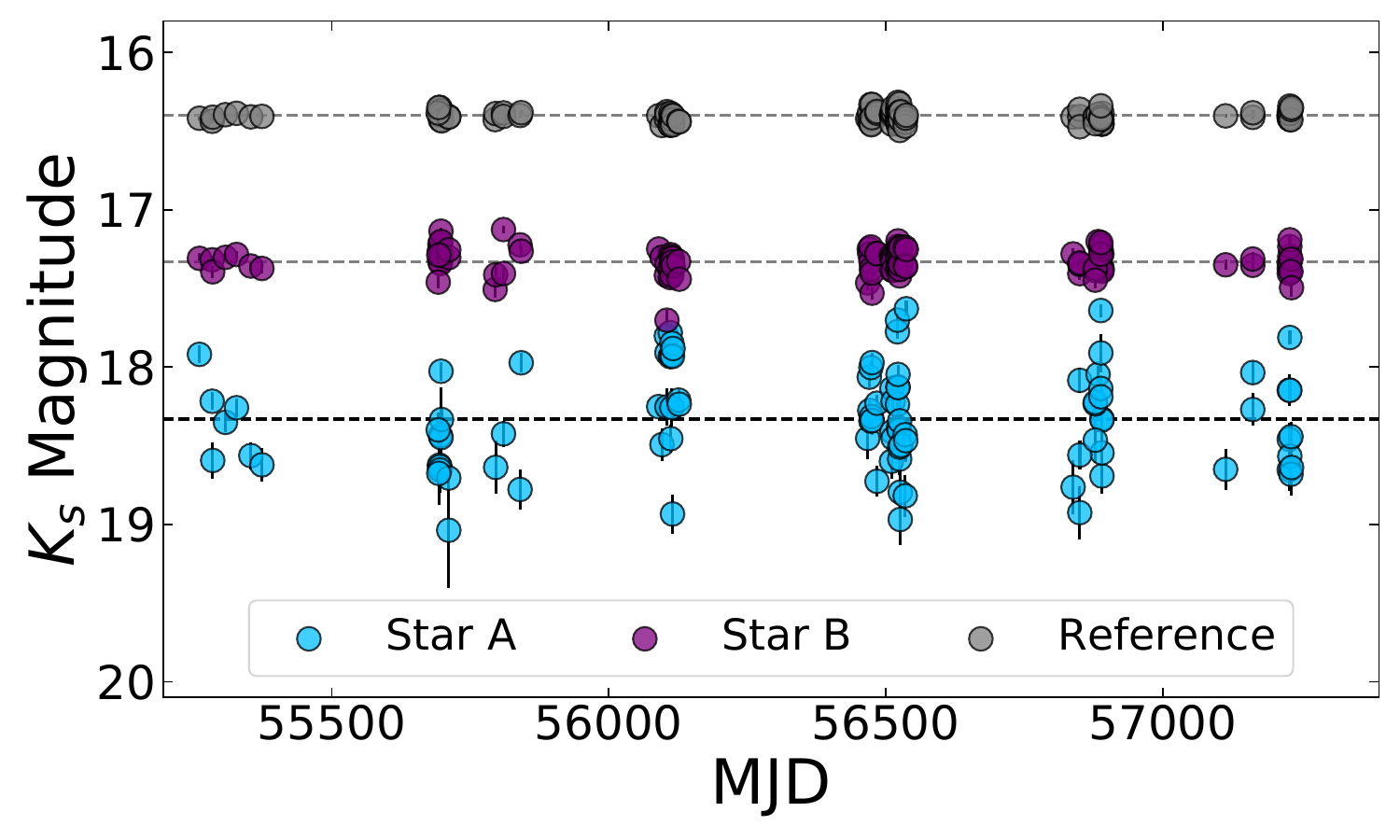}
\caption{Infrared lightcurves in the $K_s$ filter (AB magnitude) from archival VVV observations. Star A is shown in blue, Star B in purple, and a reference star is displayed in gray. The dashed lines show the median magnitude for each object. Star A displays significant variability compared both to the reference object and Star B. 
}
\label{fig: vvv_lc}
\end{figure}

We used the VVV monitoring data to build a lightcurve in the $K_s$-band to search for infrared variability. We first retrieved all data covering the position of the source from the Vista Science Archive. 
We performed point spread function (PSF) photometry on 104 VVV images, each in the $K_s$-band, using a combination of \texttt{SExtractor} \citep{Bertin1996} and \texttt{PSFEx} \citep{Bertin2011}. The photometry was calibrated using the VVV Catalog, and we applied standard Vega to AB magnitude offsets. We utilized the same 10 reference stars to produce PSF models in all images. The results are presented in Figure \ref{fig: vvv_lc}. 

\begin{figure}
\centering
\includegraphics[width=\columnwidth]{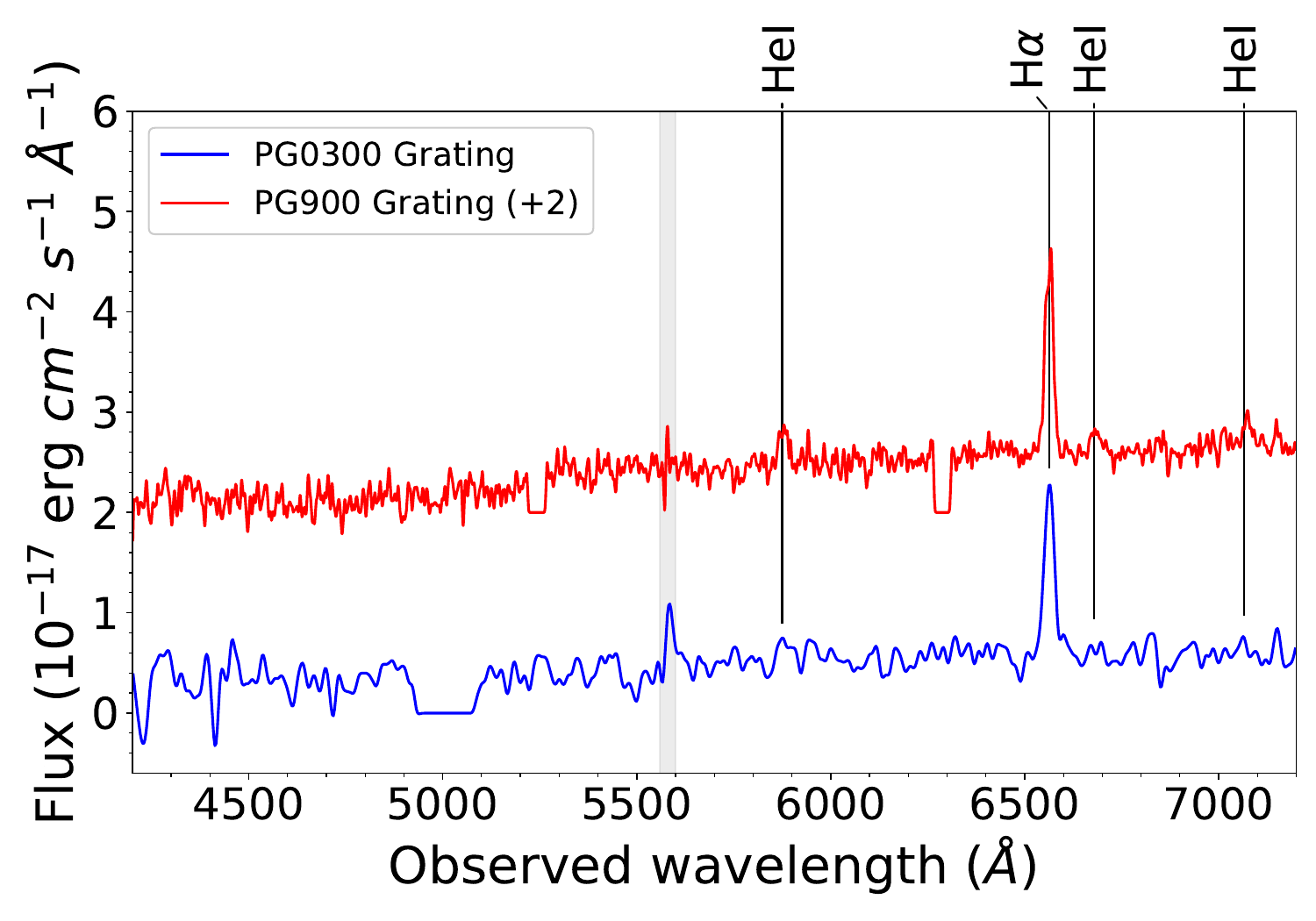}
\caption{SALT/RSS spectra of Star A obtained on 2022 March 28 (PG0300; lower resolution) and 2022 April 28 (PG900; higher resolution). The spectra are not corrected for extinction, and have been smoothed for display purposes using a Savitzky-Golay filter. The light gray line marks a telluric region, and the chip gaps are visible at 5250 and 6300 \AA\, in the red spectrum (top) and 5000 \AA\, in the blue spectrum (bottom). 
}
\label{fig: spectra}
\end{figure}

\subsubsection{Southern African Large Telescope (SALT)}
\label{sec: salt}

Given the indication of $H\alpha$ from VPHAS, we targeted the counterpart with the Robert Stobie Spectrograph \citep[RSS;][]{Burgh2003,Kobulnicky2003,Smith2006} mounted on the 11-m SALT \citep{Buckley2006} on 2022 March 28 starting at 00:17:29 UT. A spectrum was acquired using the PG0300 grating, yielding a central wavelength of $6210$\,\AA, with a total exposure of 1500 s. A second epoch of spectroscopy was acquired on 2022 April 28 starting at 21:57:57 UT using the higher resolution PG900 grating with a series of two exposures of 1200 s each. 
The spectrum was reduced using the \texttt{PySALT} package \citep{Crawford2010} and flux calibrated using the spectrophotometric standard star HILT600. We further scaled the spectrum to match the VPHAS photometry, and combined the two exposures. Our spectra display a clear emission line consistent with H$\alpha$ (Figure \ref{fig: spectra}), and tentative lines of HeI. There are no emission lines detected at $<5500$ \AA. This is likely due to rapidly decreasing SNR given the source faintness at these wavelengths: $g>22.8$ mag compared to $r\approx21$ mag.

\section{Results} 
\label{sec: results}

\subsection{Timing Analysis}
\label{sec: timing}

\begin{figure*}
\centering
\includegraphics[width=1.7\columnwidth]{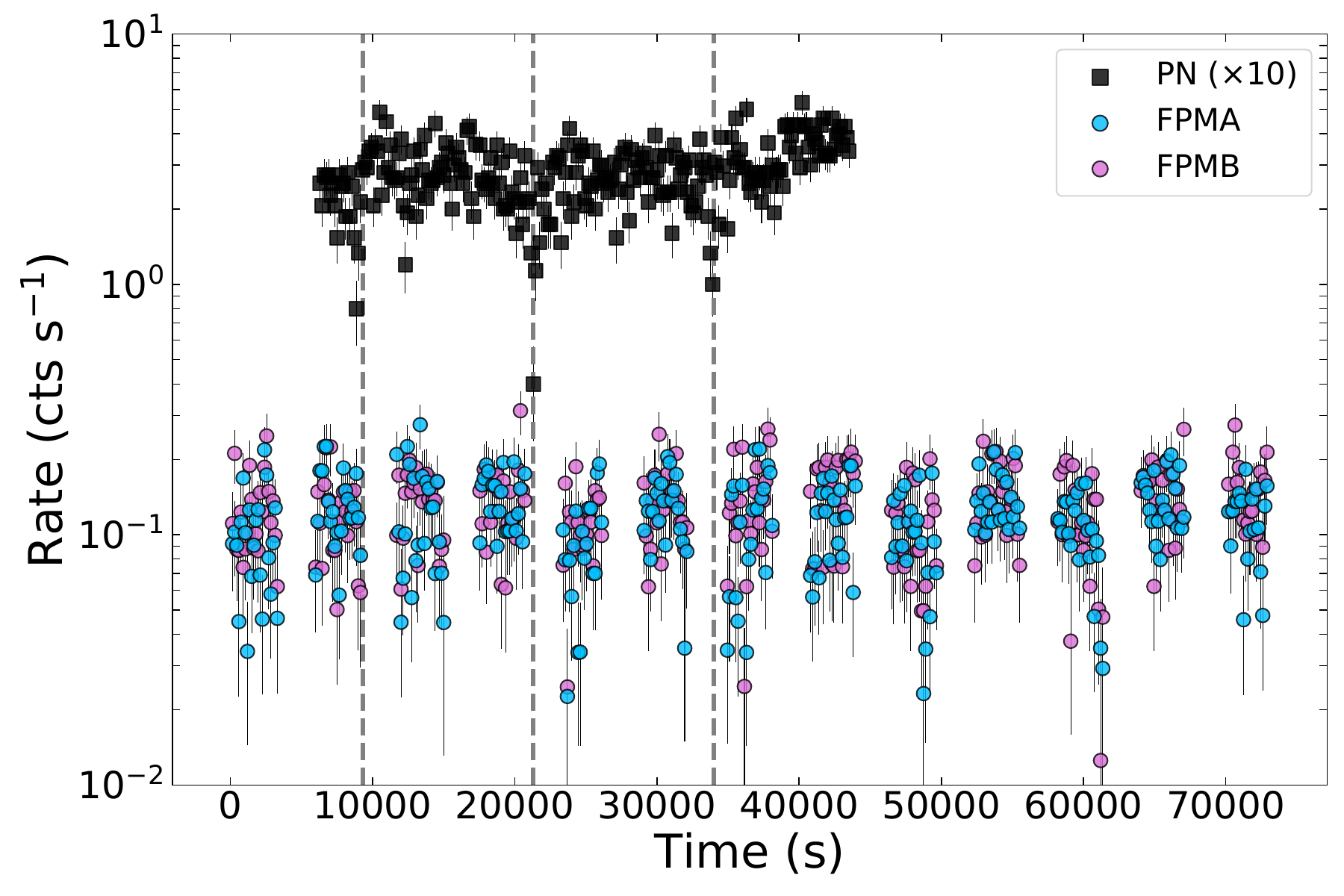}
\caption{X-ray lightcurves from \textit{NuSTAR} (150 s bins; $3-79$ keV) and \textit{XMM-Newton} (150 s bins; $0.3-10$ keV) observations. The dashed gray vertical lines mark the time of minima observed in the \textit{XMM-Newton}/PN lightcurve. The lines are each separated by $12,000$ s. 
}
\label{fig: xraylc}
\end{figure*}

\begin{figure}
\centering
\includegraphics[width=\columnwidth]{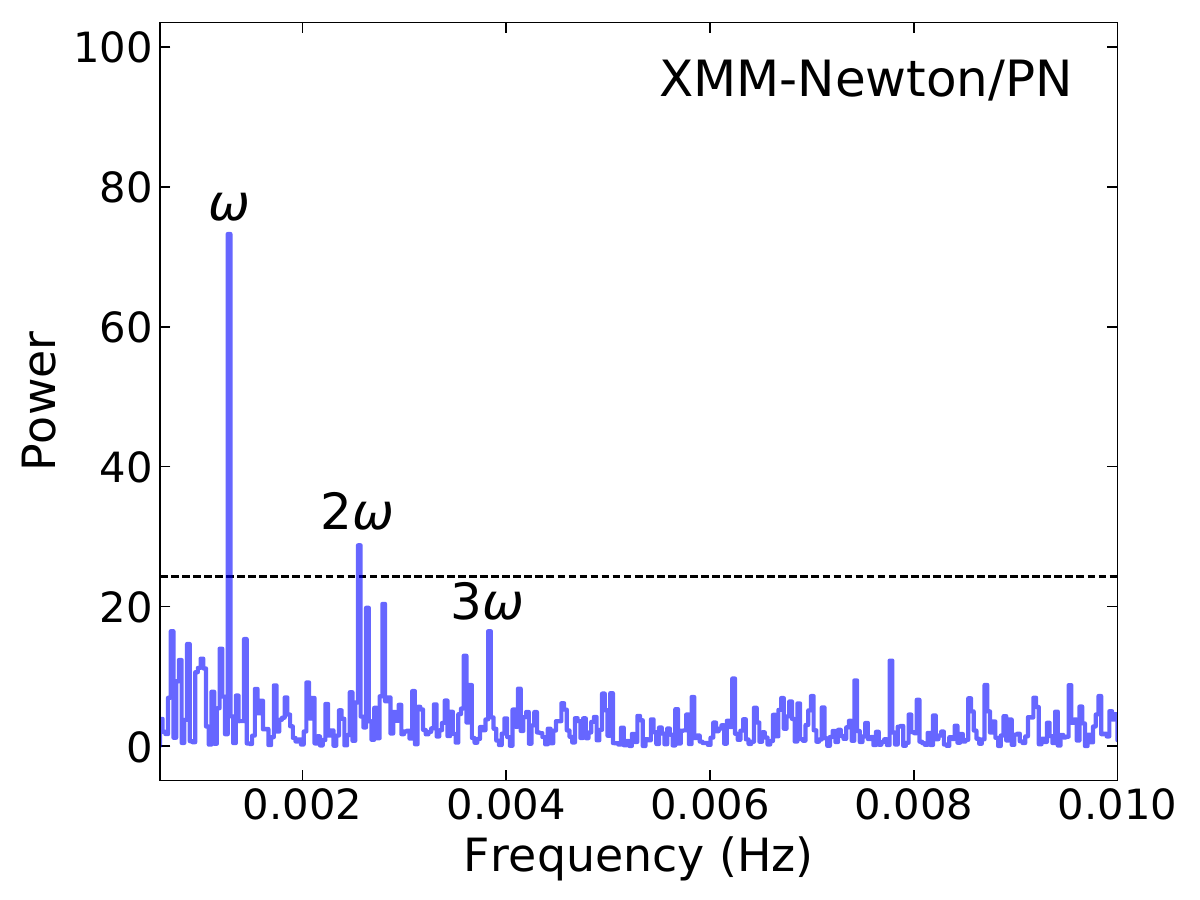}
\caption{Periodogram of our \textit{XMM-Newton}/PN data. A significant signal exists at $1.2753992\times10^{-3}$ Hz, with multiple harmonics also detected. The dotted black line marks a $3\sigma$ significance.
}
\label{fig: periodogram}
\end{figure}


The X-ray lightcurves from our simultaneous \textit{NuSTAR} and \textit{XMM-Newton} ToO are displayed in Figure \ref{fig: xraylc}. 
We searched the \textit{XMM-Newton} data for coherent and quasi-coherent signals using periodogram analysis \citep{Huppenkothen2019}. Averaging periodograms from all three detectors (PN/MOS1/MOS2) and comparing the highest outlier in the averaged periodogram against a null hypothesis of pure detector noise, we reject the null hypothesis at a significance level of $p = 2.6 \times 10^{-26}$ for the highest outlier at $1.2753992\times10^{-3}$ Hz, corresponding to 784 s, an extremely strong rejection of the null hypothesis. The signal is significantly detected at a high confidence in individual detectors (MOS1: $p = 2.3 \times 10^{-8}$; PN: $p = 1.10 \times 10^{-12}$). The threshold for a $3\sigma$ detection, corrected for the number of frequencies searched, is $p < 1.4 \times 10^{-6}$. We note a clear peak at the first harmonic (exactly twice the frequency) of the detected signal (see Figure \ref{fig: periodogram}). 

However, the periodogram also reveals the presence of stochastic variability (red noise) in the data, which might contribute significantly to the power observed at the relevant frequencies. We thus follow the procedure of \citet{Vaughan2010} to estimate the posterior predictive probability for the presence of a periodic  signal in the presence of red noise, using a simple power law model and a constant as our baseline model. We employ Markov Chain Monte Carlo Sampling with wide, uninformative priors to draw from the posterior for the model parameters 
and generate simulated periodograms in order to sample the posterior predictive distribution for the highest outlier. We find a posterior predictive $p$-value for the highest outlier in the averaged periodogram across the three detectors of $p=0.0011$ (corrected for the number of trials). Thus, the significance of the signal is at the $>3\sigma$ level in the \textit{XMM-Newton} data. 

However, no significant coherent pulsation is detected in a simultaneous \textit{NuSTAR} observation using the same method. We find that this is likely due to the data gaps caused by the low earth orbit of the satellite. Similarly, we do not find the signal in our \textit{NICER} observations, likely due to the faintness of the source and the large gaps in the data.

\begin{figure}
\centering
\includegraphics[width=\columnwidth]{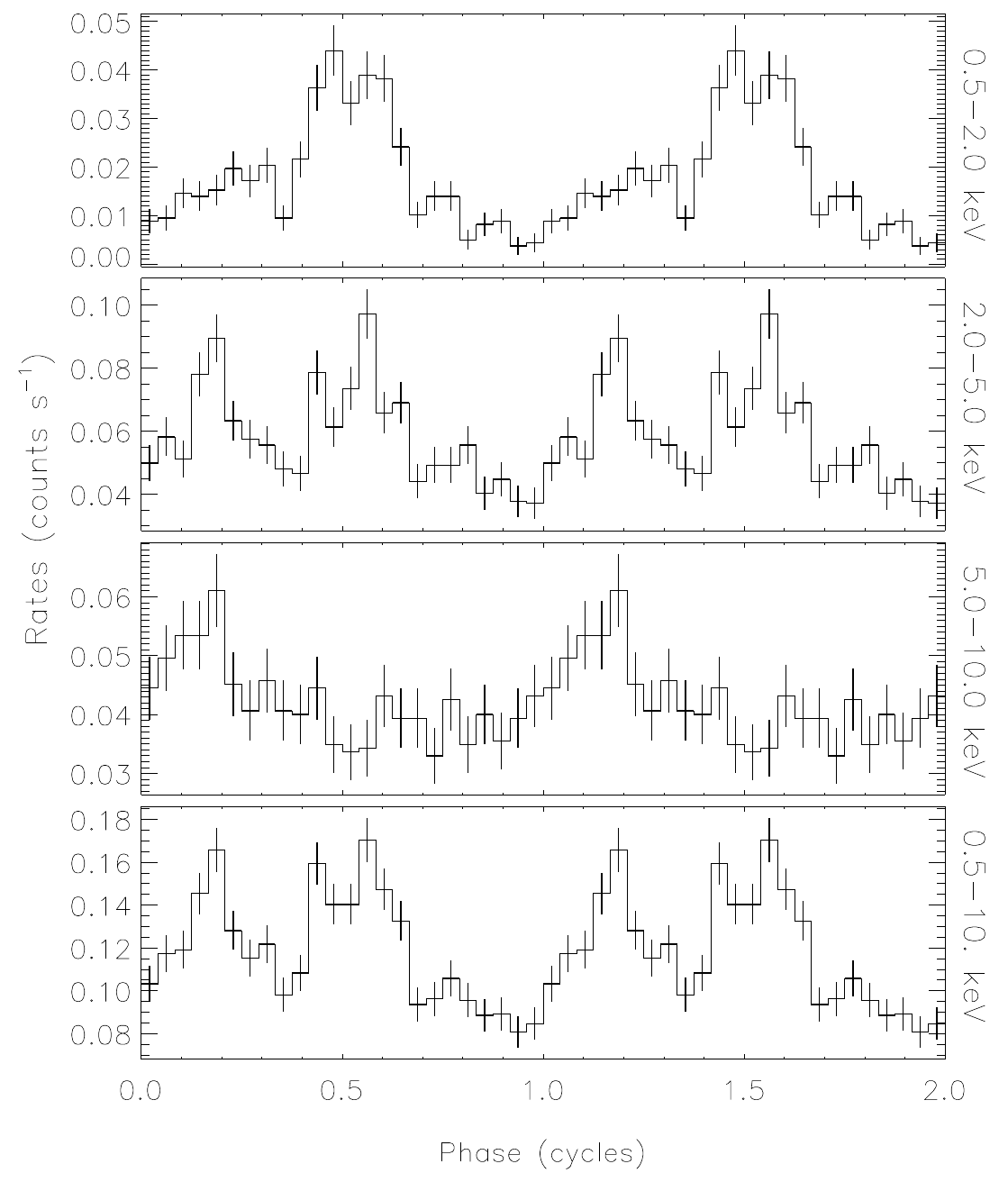}
\caption{Phase folded lightcurve from \textit{XMM-Newton} in multiple energy bands ($0.5-2$, $2-5$, $5-10$, $0.5-10$ keV) using a period of 784 s ($1.2753992\times10^{-3}$ Hz).
}
\label{fig: phase_folded_lc}
\end{figure}

The phase-folded \textit{XMM-Newton} lightcurve (Figure \ref{fig: phase_folded_lc}) displays multiple peaks with a significant energy dependence. 
The first peak is seen above 2 keV, the second peak is exclusively seen below 5 keV. The energy dependence of the spin period implies that the system may have a variable accretion column or partial covering fraction. The RMS pulsed fraction decreases with energy from $59\pm5\%$ at $0.5-2$ keV to $21\pm2\%$ at between $2-5$ keV to $12\pm3\%$ at $5-10$ keV. The overall RMS pulsed fraction is $19\pm2\%$ in the $0.5-10$ keV energy range.

Despite the fact we do not find any significant signal in the \textit{NuSTAR} data, we fold the lightcurve at the 784 s period to test the presence of a signal. We obtain a $3\sigma$ upper limit to the RMS pulsed fraction of $<9\%$ in $3-10$ keV and $<8\%$ in $3-79$ keV. 
The rapidly decreasing pulsed fraction observed in the \textit{XMM-Newton} data may contribute to the difficulty in extracting the signal from the \textit{NuSTAR} data.

While we do not observe any other significant signals in our data at lower frequencies, in Figure \ref{fig: xraylc} we have marked three times of minima in the \textit{XMM-Newton}/PN lightcurve. These minima are separated by $\sim12,000$ s ($\sim3.3$ hr). This may be the orbital period of the system, but given the low significance, and the orbital gaps in the \textit{NuSTAR} data, we cannot confirm this feature. Future observations are required to determine if this is a coherent orbital period. 

\subsection{Spectral Analysis}
\label{sec: specanalysis}

\begin{figure}
\centering
\includegraphics[width=\columnwidth]{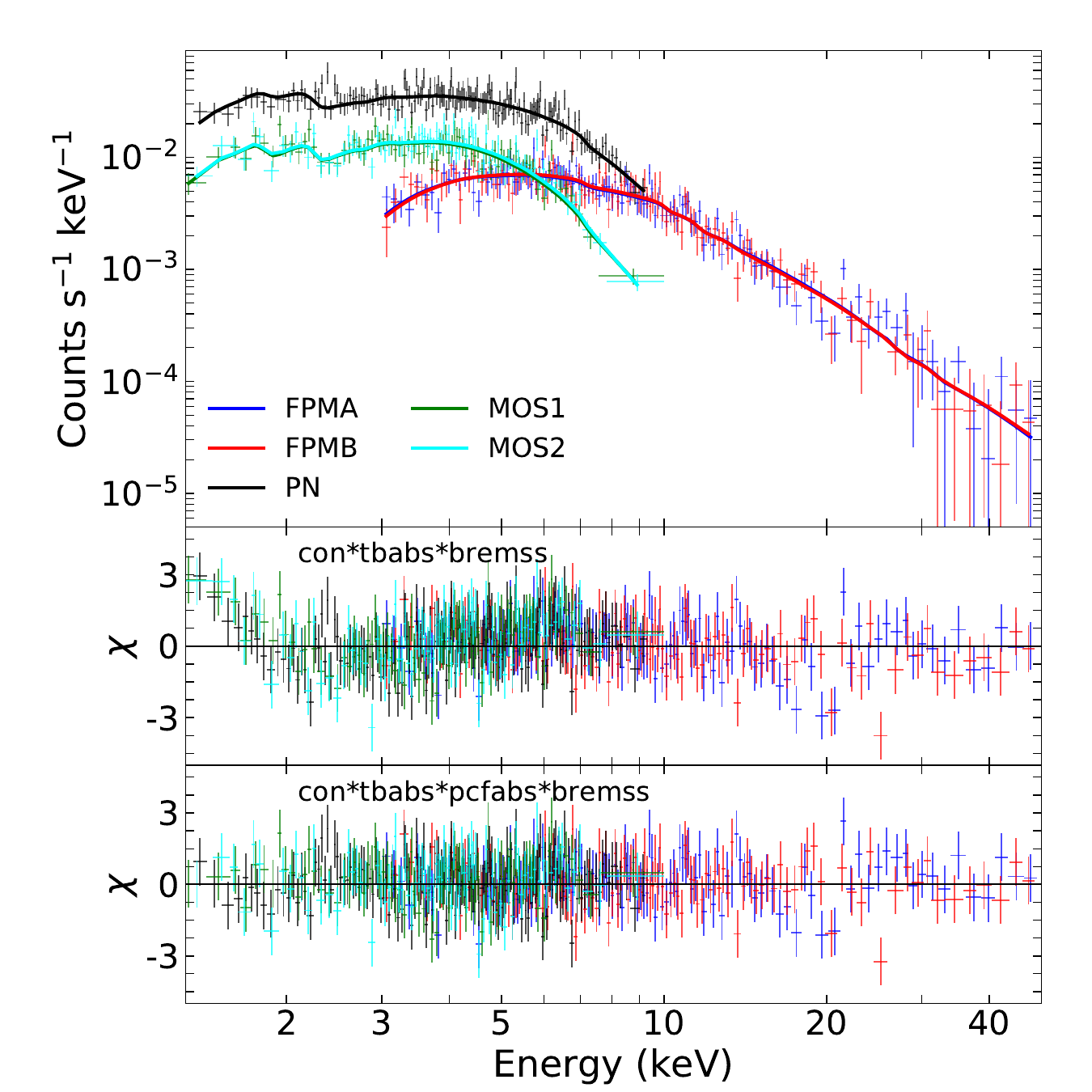}
\caption{Top panel: Best fit models to the X-ray spectra obtained simultaneously with \textit{NuSTAR} and \textit{XMM-Newton}. 
Bottom panels: fit residuals for two different spectral models with continuum emission due to thermal bremsstrahlung radiation. 
}
\label{fig: xrayspec}
\end{figure}

The time-averaged spectra from our simultaneous \texttt{XMM-Newton} (PN/MOS1/MOS2) and \texttt{NuSTAR} (FPMA/FPMB) observations appear featureless, with no clear emission or absorption lines. We jointly model these spectra using \texttt{XSPEC 12.12.0} \citep{Arnaud1996}. We applied the ISM abundance table from \citet{Wilms2000} and the photoelectric absorption cross-sections presented by  \citet{Verner1996}. The best fit model parameters were derived by minimizing the Cash statistic \citep{Cash1979}. The spectra are displayed in Figure \ref{fig: xrayspec}.

Using an absorbed power-law model (\texttt{con*tbabs*pow}), we find a hydrogen column density $N_H=(5.0\pm0.2)\times10^{22}$ cm$^{-2}$ and photon index $\Gamma=1.47\pm0.03$ with Cstat = 639 for 576 dof. This model has significant residuals at $<$\,$2$ keV, implying that a further partial covering absorber may be required. Adding an additional component (\texttt{pcfabs}) improves the fit statistic (Cstat = 518 for 574 dof) and removes the low energy residuals. In this case, we find $N_{H,\texttt{tbabs}}=(2.9\pm0.3)\times10^{22}$ cm$^{-2}$, $N_{H,\texttt{pcfabs}}=(13.5^{+2.2}_{-1.9})\times10^{22}$ cm$^{-2}$, partial covering fraction of $0.70\pm0.04$, and photon index $\Gamma=1.74\pm0.05$. The absorbed X-ray flux at the time of these observations is $\approx(1.85-2.08)\times10^{-12}$ erg cm$^{-2}$ s$^{-1}$ and the unabsorbed ($0.3-10$ keV) flux is $\approx(3-4)\times10^{-12}$ erg cm$^{-2}$ s$^{-1}$.

We also tested an absorbed cutoff power-law (\texttt{con*tbabs*cutoffpl}) and a Comptonization model \texttt{CompTT} in \texttt{XSPEC} (\texttt{con*tbabs*CompTT}; \citealt{Titarchuk1994}). The Comptonization model is described by the temperature of seed photons $kT_0$ that are then Comptonized by a plasma of temperature $kT_1$ with optical depth $\tau$. For the cutoff power-law, we obtain $N_H=(3.3\pm0.2)\times10^{22}$ cm$^{-2}$ and photon index $\Gamma=0.66\pm0.05$ with cutoff energy $E_\textrm{cut}=11\pm2$ keV.
Whereas, for the Comptonization model we derive $kT_0=1.2\pm0.04$ keV and $kT_1\approx 21.5$ keV with optical depth $\tau\approx2.2$ for $N_H=(0.90\pm0.15)\times10^{22}$ cm$^{-2}$ (Cstat = 517 for 574 dof). The parameters $kT_1$ and $\tau$ are largely unconstrained by the model. These two models provide a similar description of the data, and do not require the addition of another absorption component.

We also tested a thermal bremsstrahlung model (\texttt{bremss}) and found that it  provides a similarly good description for a temperature $kT\approx45$ keV (Cstat = 599 for 576 dof). In this case, however, we find that the addition of a partial covering fraction is required to remove the low energy residuals (Figure \ref{fig: xrayspec}). After adding this component (\texttt{pcfabs}), we derive $N_{H,\texttt{tbabs}}=(2.2\pm0.4)\times10^{22}$ cm$^{-2}$, $N_{H,\texttt{pcfabs}}=(9.3^{+1.9}_{-1.4})\times10^{22}$ cm$^{-2}$, partial covering fraction of $0.65\pm0.08$, and temperature $kT=29.5\pm3.5$ keV with Cstat = 521 for 574 dof. 
Motivated by the bremsstrahlung solution, we considered also an \texttt{mkcflow} model which correctly accounts for both a bremsstrahlung continuum and line cooling with the inclusion of multiple plasma lines (He-like at 6.7 keV and H-like at 6.97 keV). We find that this model is largely unconstrained by our data, but does also require a \texttt{pcfabs} component. 
While there is no clear indication of a preferred model based on our analyses, we can conclude that the source is a luminous hard X-ray emitter with an absorbing column that is likely in excess of the Galactic value of $N_H=1.6\times10^{22}$ cm$^{-2}$ \citep{Willingale2013}. 

We performed an additional check to search for the significance of a 6.4 keV iron fluorescence line in the data. We added an additional Gaussian emission component (\texttt{gauss}) with fixed energy 6.4 keV and fixed width 0.1 keV to our best fit model. We derive a $3\sigma$ upper limit to the equivalent width (EW) of this component of $<166$ eV. Therefore, while our spectra appear featureless, we cannot exclude additional emission components using these observations alone.

As a last step, we investigated the phase-resolved spectra of J1708. We used the \texttt{SAS} software to produce phase-resolved spectra and responses for phases with a pulse (on-pulse) and without a pulse (off-pulse), see Figure \ref{fig: phase_folded_lc}. We define the on-pulse phases as between $0.1-0.3$ and $0.4-0.8$, and the off-pulse phases are all others. 
The two spectra vary only marginally below $3$ keV, which we attribute to a slight change in partial covering fraction with phase. We derive a higher partial covering fraction during the off-pulse phases. 
A change in the partial covering fraction is typically how X-ray variability in IPs displays itself, and this is not unexpected.

\subsection{Multi-wavelength counterpart}
\label{sec: counterpart}

Using archival observations (\S \ref{sec: archival}) from \textit{Gaia}, VVV, and VPHAS, we identified Star A (Figure \ref{fig: counterpart}) as the likely counterpart to J1708, given its coincidence to our \textit{Chandra} localization (\S \ref{sec: chandra}). Follow-up spectroscopy with SALT (\S \ref{sec: salt}) revealed multiple emission features (Figure \ref{fig: spectra}) including a clear detection of H$\alpha$ and evidence for three HeI lines. 

In Figure \ref{fig: vvv_lc}, we present the $K_s$-band lightcurve of Star A compared to both Star B and a nearby reference star. In comparison to the reference objects, Star A displays significant variability of $\pm0.7$ mag (this is a range and not a standard deviation) around a median value of $K_s\approx18.3$ AB mag. Thus the variability amplitude is $\approx$ 1.4 mag, compared to $\approx$ 0.5 mag for a nearby star of similar faintness, due likely to variable seeing and the complex background in the crowded field. We searched the photometry using a Lomb-Scargle analysis, but, due to the sparse sampling of the data, no significant periodic signals were recovered.

\begin{figure}
\centering
\includegraphics[width=\columnwidth]{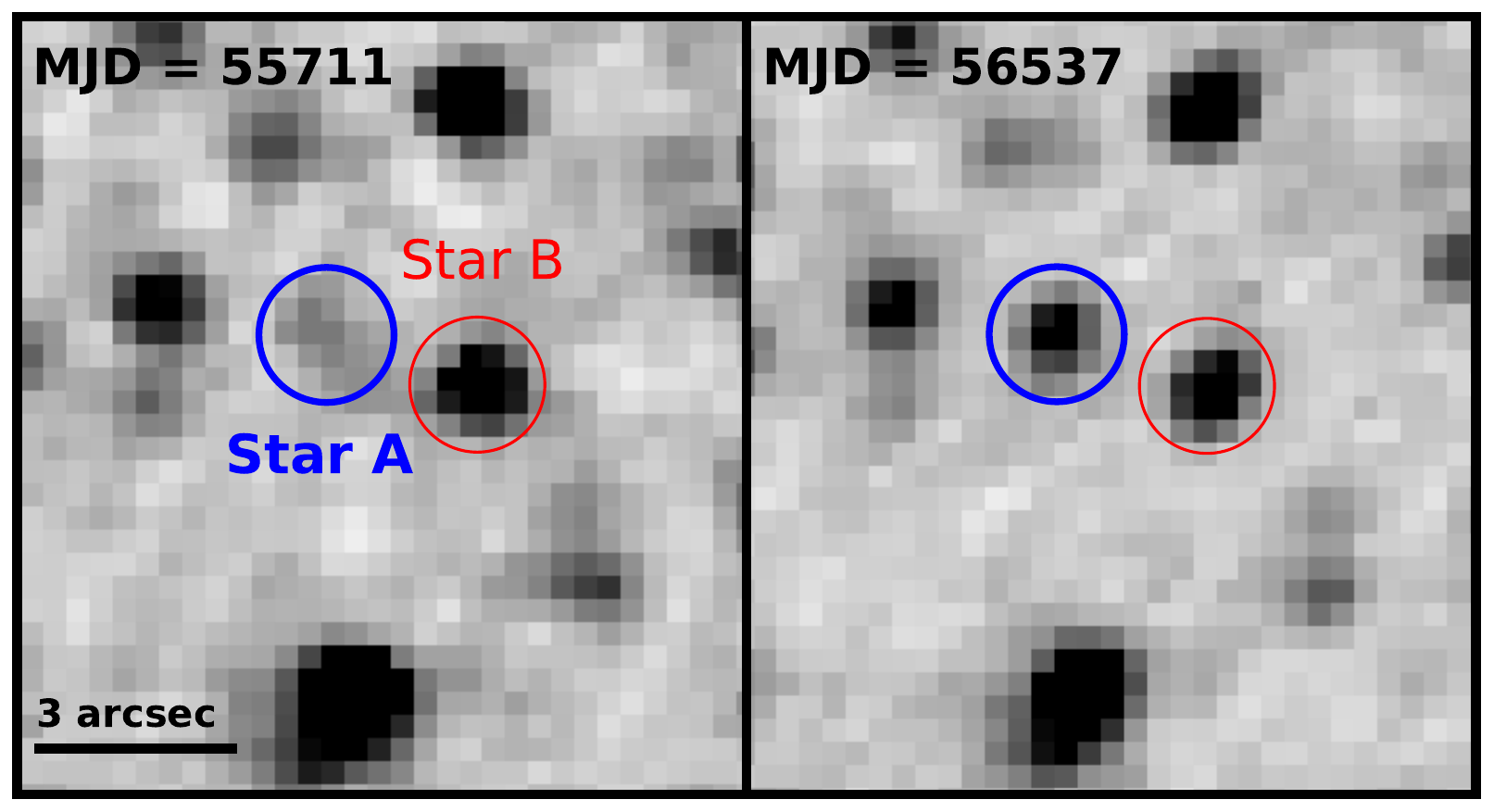}
\caption{VVV $K_s$-band single epoch images of the infrared counterpart to J1708, which is labeled in blue as Star A (left circle). The nearby Star B is also labeled for comparison (right circle). The blue and red circles show the location of both stars, respectively. The left image shows the faintest observation of Star A, and the right image displays the brightest epoch. 
North is up and East is to the left. 
}
\label{fig: varfig}
\end{figure}

As additional evidence, we display a finding chart showing a comparison between the faintest and brightest detections of Star A from the VVV data (Figure \ref{fig: varfig}). On MJD 55711 (left image) we measure a brightness of $K_s=19.0\pm0.4$ AB mag and on MJD 56537 (right image) we derive $K_s=17.62\pm0.05$ AB mag. 
Both epochs have seeing of 1.1\arcsec and 1.0\arcsec, respectively, which has no significant impact on the image comparison. Comparing the photometry from these epochs, the variability amplitude is $1.4\pm0.4$ mag.  Based on our photometry, we suggest that Star A is significantly variable.

\begin{figure}
\centering
\includegraphics[width=\columnwidth]{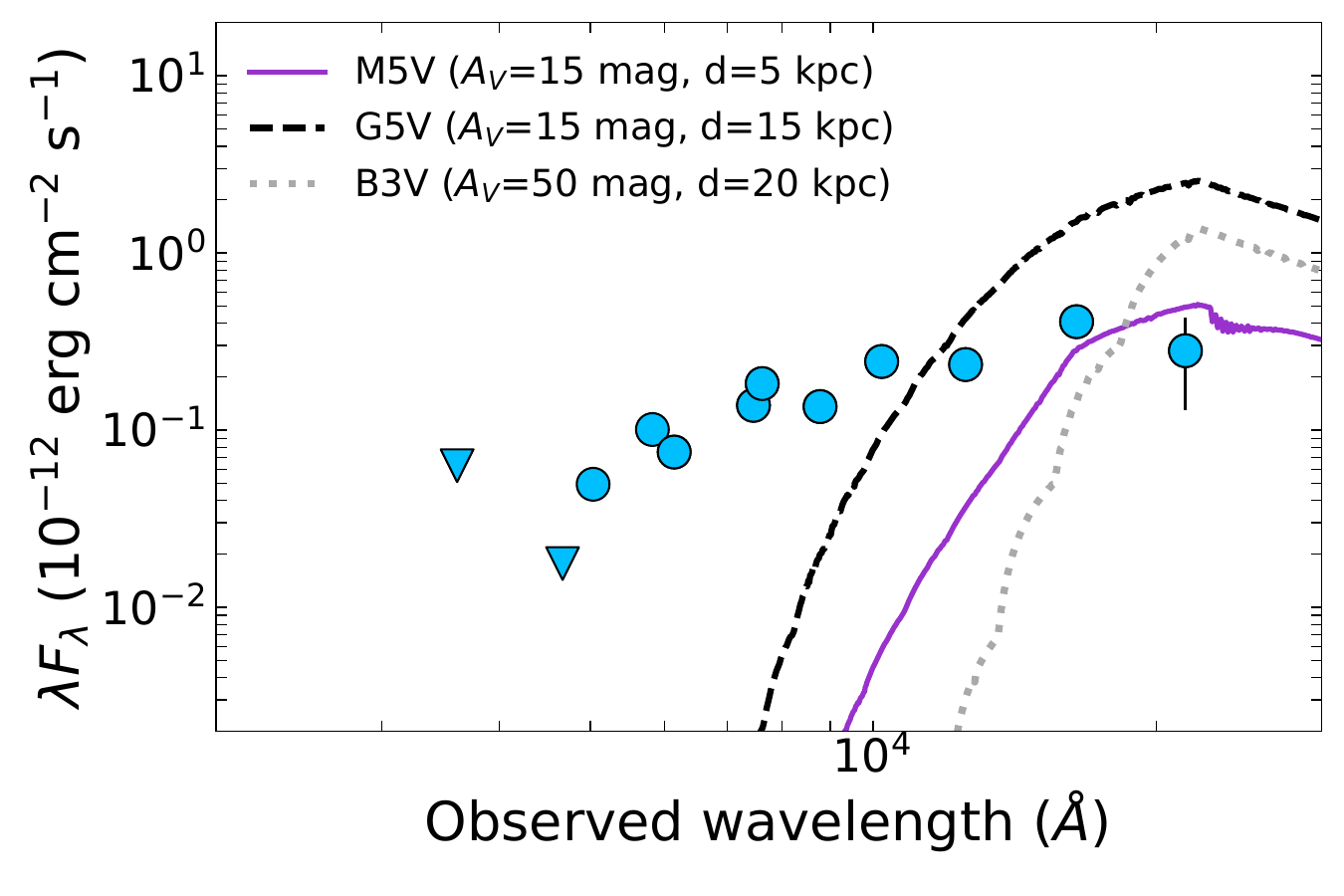}
\caption{Archival optical and infrared SED of the multi-wavelength counterpart (Star A) compared to different stellar atmosphere models \citep{Kurucz1993}. Downward triangles represent $3\sigma$ upper limits. The data are not corrected for the unknown extinction. The uncertainty in the $K_s$-band photometry reflects the $\pm0.7$ mag variability in the VVV data.
}
\label{fig: counterpartSED}
\end{figure}

The emission lines, and significant infrared variability, imply that accretion is ongoing in the system, and that the dominant source of emission observed in our SED (Figure \ref{fig: counterpartSED}) is due to an accretion disk or column. Therefore, the potential donor star contribution is likely overshadowed, and for this reason we do not attempt to classify a stellar type using the observed archival photometry.

However, for comparison, we display multiple absorbed stellar SEDs \citep{Kurucz1993} to set tentative constraints to the type of donor star (Figure \ref{fig: counterpartSED}). 
As there is no \textit{Gaia} parallax measurement, the distance to the source is unconstrained, but likely to be at least a few kpc. 
Utilizing the $N_H=(0.9-3.3)\times10^{22}$ derived from the X-ray spectra (\texttt{comptt} and \texttt{cutoffpl} models), and assuming the \citet{Guver2009} relation between optical extinction and hydrogen column density we find $A_V=4-15$ mag. 
This is significantly less than the full Galactic value along the line of sight ($A_V=84$ mag; \citealt{Schlafly2011}), although we caution that the \citet{Guver2009} relation may not apply here given the possibility of additional extinction and absorption intrinsic to the system. Furthermore, comparing this to Galactic extinction maps \citep{Amores2021} suggests an approximate distance of between $3-5$ kpc for Galactic $l,b$ = 346.165251 deg, -0.015251 deg, but this is also likely a highly uncertain estimate. In any case, we determine that it is likely safe to assume that the distance to J1708 is at least a few kpc.

We find that, due to the counterpart's faintness ($r=20.93\pm0.10$ AB mag), the SED disfavors a high-mass donor star, and instead suggests that low-mass donor stars (e.g., M dwarfs) are more consistent with the data (Figure \ref{fig: counterpartSED}). High-mass donor stars require increasingly further distances and larger values of extinction to be consistent with the data, even ignoring the contribution from accretion.

Furthermore, by inspecting archival \textit{Spitzer} images from the Galactic Legacy Infrared Mid-Plane Survey Extraordinaire \citep[GLIMPSE;][]{Benjamin2003} we confirm that no source is detected at the position of Star A. As \textit{Spitzer} is largely unaffected by Galactic extinction, we consider this a robust constraint against a high-mass counterpart, which would require a distance $>20$ kpc to not be  detected above 3.6 $\mu$m (see, e.g., \citealt{OConnor2021}).  

\section{Discussion}
\label{sec: discussion}



Due to our detection of a hard X-ray spectrum and H$\alpha$ emission feature, we consider it likely that J1708 is an accreting binary. In what follows, we consider multiple interpretations of the source as either an HMXB, LMXB, or magnetic CV.

\subsection{Interpretation as a High-mass X-ray Binary (HMXB)}

The coherent periodicity of 784 s is consistent with the observed spin periods of NSs in Be X-ray Binary (BeXRB) systems \citep{Corbet1984,Corbet2017}. We note, however, that a decreasing pulsed fraction with energy is atypical for a BeXRB. Moreover, in order for a massive stellar companion to be consistent with the observed SED (Figure \ref{fig: counterpartSED}), we would require that the source is both highly extinct ($A_V\gtrsim50$ mag) and at a large distance ($\gtrsim20$ kpc). The lack of an infrared detection in archival \textit{Spitzer} observations of the field is strange for a BeXRB. Assuming a distance of 20 kpc, the persistent X-ray luminosity is $\approx2\times10^{35}$ erg s$^{-1}$, which would imply an intermediate luminosity BeXRB \citep{Reig1999}. We disfavor other classes of HMXBs, such as supergiant fast X-ray transients, due to the lack of observed flaring or outbursts in archival X-ray observations.

\subsection{Interpretation as a Low-mass X-ray Binary (LMXB)}

The observed spin periods of NSs in LMXB systems are on the order of milliseconds (e.g., 400 Hz; \citealt{Gusakov2014}). This is significantly shorter than our observed 784 s (13 min) signal.  In addition, coherent pulsations and a high pulsed fraction are atypical for LMXBs, which tend only to show millisecond pulsations transiently during outbursts \citep{Dib2005,Patruno2018}. Assuming a tentative distance of $3-5$ kpc, the source has a $0.3-10$ keV X-ray luminosity of $10^{33-34}$ erg s$^{-1}$, which is at the extremely faint end of the LMXB luminosity function \citep{Revnivtsev2008lmxb,Revnivtse2011}. 

On the other hand, a rare sub-class of LMXB display short orbital periods ($<80$ min), e.g., 4U 1820-30 \citep{Priedhorsky1984,Stella1987}, 4U 1543-624 \citep{Wang2004}, and 2S 0918-549 \citep{intZand2005}. 
The detection of an H$\alpha$ emission line in our optical spectra implies a hydrogen-rich donor star, which is inconsistent with such a short orbital period \citep[e.g.,][]{Nelemans2010}.
As only a handful of these sources are known, we consider the interpretation of an ultra-short period LMXB unlikely, due also to the lack of X-ray bursts or variability.

\subsection{Interpretation as a Cataclysmic Variable (CV)}

Here, we consider the possibility that J1708 is a magnetic CV based on the following evidence: 
\textit{i}) a hard X-ray spectrum, 
\textit{ii}) a decreasing pulsed fraction with energy, 
\textit{iii}) a likely low-mass donor star, and
\textit{iv}) multi-wavelength variability. Moreover, the 784 s signal is a typical spin period for WDs in CVs. 

The X-ray flux of J1708 has remained largely constant (within errors) since its first discovery in 2012. The absorbed X-ray flux ($0.3-10$ keV) in all of our soft X-ray observations (\textit{Swift}, \textit{XMM-Newton}, \textit{NICER}, and \textit{Chandra}) is $\approx2\times10^{-12}$ erg cm$^{-2}$ s$^{-1}$. 
The X-ray spectra favor a partially covered thermal bremsstrahlung model, which is typically observed in IP CVs \citep[e.g.,][]{Tomsick2016,Mukai2017,Shaw2020,Hare2021,Gorgone2021}. 
Our data quality is not high enough to constrain a reflection component, or to detect emission features. Typical magnetic CVs display a 6.4 keV fluorescent, 6.7 keV He-like, and 6.97 keV H-like Fe K lines. Using our X-ray spectra, we set an upper limit of $<166$ eV to the EW of the Fe fluorescence line at 6.4 keV. This value is consistent with the  measurements (EW$\approx93$ eV) from other magnetic CVs \citep{Romanus2015}, see also \citet{Ezuka1999}, and does not rule out such a classification.

The source X-ray luminosity of $10^{33-34}$ erg s$^{-1}$, assuming a distance of $3-5$ kpc, is consistent with luminous IPs detected with \textit{INTEGRAL} \citep{Revnivtsev2008,deMartino2020}. Based on the hard X-ray spectrum, decreasing pulsed fraction, and high X-ray luminosity, we suggest that J1708 is an IP CV. Polar CVs instead show a lower X-ray luminosity and are synchronous, such that the orbital and spin periods are locked. The 13 min periodicity, therefore, strongly disfavors a polar CV interpretation. 

Typically, the X-ray light curves of IPs are dominated by periodic modulations (on spin, orbital, and occasional sideband periods) but are  otherwise flat within a given X-ray observation \citep{Taylor1997,deMartino2001,Hellier2002}. Figure \ref{fig: xraylc} shows the X-ray lightcurves from simultaneous \textit{XMM-Newton} and \textit{NuSTAR} observations. While the source variability is not extreme, it is slightly more than expected for an IP. This does not conclusively exclude our interpretation of the source as an IP, but requires further investigation. We note that there is a hint of a repeated structure with period $\sim3.3$ hr in the \textit{XMM-Newton} lightcurve (Figure \ref{fig: xraylc}), which would explain the observed X-ray variability as the orbital period of the binary. Confirmation of the orbital period would solidify our interpretation. 

An alternative possibility is that J1708 is an AM Canum Venaticorum (AM CVn) system, such as the 10 min helium binary ES Ceti \citep{Downes1993}. These short period binaries are known to have hydrogen-deficient donor stars \citep{Nelemans2010}. 
Therefore, the optical spectra of these systems are dominated by emission lines from the HeII Pickering series \citep{Bakowska2021}, which mimics the Balmer series (albeit with additional lines). However, as we find no evidence for HeII $\lambda$5411 in our spectra, we cannot confirm this interpretation. Under the assumption that the bright emission feature is H$\alpha$, we can exclude a short period binary. 

\subsubsection{Origin of the pulse profile}

While the decreasing pulsed fraction is typical of CVs \citep{Mukai2017}, the multi-peaked, energy-dependent pulse profile is not a common feature. The predominant cause of the X-ray spin modulation is thought to be absorption in the pre-shock flow from the so-called ``accretion curtains'' \citep{Norton1989}, although occultation of the emission region can also contribute \citep{King1984}. Generally IPs display X-ray spin modulations that can be approximated by a single sinusoidal profile \citep[e.g.,][]{Norton1989,Bernardini2012}. However, a handful of IPs display double sinusoidal profiles with two peaks per cycle \citep{Norton1999}, e.g., FO Aqr \citep{Norton1992} and PQ Gem \citep{Duck1994}.

In fact, the spin profile of PQ Gem \citep{Duck1994} appears similar to J1708 above 0.6 keV, where the X-ray emission is dominated by the shock. At energies $<0.5$ keV, the X-ray emission from PQ Gem is dominated by an optically thick, blackbody-like component emitted from the WD surface, which is not observed in J708. 

The complication, in the case of J1708, is instead the energy dependence of the double peaked pulse profile. The light curve appears single peaked at both $0.5-2$ keV and $5-10$ keV, but a clear secondary peak arises at $2-5$ keV. The peaks display themselves as a primary (soft) and secondary (hard) pulse (see Figure \ref{fig: phase_folded_lc}). During phases $0.4-0.7$, we observe the primary peak at low and medium energies, while we see a narrower, secondary peak around phase $\sim$\,$0.2$ at medium and high energies. 

While our phase-resolved spectroscopy revealed only marginal deviations in the X-ray spectra of the on-pulse versus off-pulse emission, we did derive a higher partial covering fraction during the off-pulse phase. This suggests that the profile is due to phase-dependent absorption \citep{Bernardini2012}, as opposed to occultation, which cannot explain the energy dependence of the pulses. 

However, the complexity of the pulse profile is further exacerbated by the equivalent pulse height of the primary and secondary peaks. As the main cause for the pulse profiles in IPs is absorption, this means that along the line of sight corresponding to each peak the value of the hydrogen column density should be lower (as we state above). However, if the main cause of the energy dependence was varying absorption along the line of sight corresponding to the primary and secondary peaks, it is difficult to explain the lack of a secondary peak at $0.5-2$ keV while maintaining a similar pulse height in $2-5$ keV. This would suggest that along the line of sight corresponding to the secondary peak, the X-ray spectrum would either have to change significantly compared to the baseline or either be more absorbed, but with a higher overall flux output along that specific line of sight. An extra absorption region can create additional features in the spin-folded light curves at low energies, but cannot make an existing (at higher energies) feature disappear. This is not easily formulated in the standard IP picture \citep[e.g.,][]{Norton1992}. 

Moreover, the width of the primary and secondary peak differs. The primary (soft) peak is wide, while the secondary (hard) peak is narrower. The difference in peak widths could indicate a geometric effect, but the exact mechanism is unclear. Therefore, we cannot easily explain the pulse profile of the source in the standard IP picture. However, the other characteristics of J1708 lead us to classify it as a candidate IP.

\subsubsection{Infrared variability}

We have uncovered a significant infrared variability with an amplitude of nearly 1.5 magnitudes between minimum and maximum brightness (Figure \ref{fig: varfig}). The variability does not clearly display itself as high-state versus low-state variability, and instead is likely due to the sparse sampling of the spin period \citep{Potter1997,Potter2012} and, possibly, orbital modulation \citep{Zuckerman1992}. The potentially low magnetic field strength of the WD is likely to lead to cyclotron emission in the IR \citep{West1987,Harrison2007}, which may contribute to the observed variability. Further optical and infrared observations with a finer time sampling are required to determine the exact nature of the variability and how it relates to the observed spin period. 



\section{Conclusions}
\label{sec: conclusions}

Our multi-wavelength follow-up campaign of J1708 identified the likely spin period of a compact object (784 s), while the H$\alpha$ and HeI lines from optical spectroscopy confirmed the system is an accreting binary. Due to the faintness of the optical/infrared counterpart, we can disfavor a massive stellar companion, such as in an HMXB, despite the spin period falling within the known distribution of NS spin periods in such systems. 
Furthermore, a 784 s spin period is incredibly slow for a NS in an LMXB, and we exclude this scenario. 
Based on these considerations, we suggest J1708 as a candidate IP CV. 

We note that further investigation of J1708 is needed to confirm the tentative $\sim$3.3 hr orbital period and solidify our classification. High speed optical or infrared photometry is highly desirable.  Additional optical spectroscopy to solidify the spectral features (e.g., HeI) and search for lines at $<5500$ \AA\, (e.g., H$\beta$, HeII $\lambda$5411, HeI $\lambda$4686) would provide valuable information to aid in the interpretation. 

J1708 is the third X-ray selected magnetic CV identified by the \textit{Swift} DGPS \citep[see also][]{Gorgone2021,OConnor2023polar}. 
In the $0.3-10$ keV energy range, the DGPS is complete to X-ray luminosities of  $L_X$\,$\sim$\,$10^{31}$ to $10^{33}$ erg s$^{-1}$ out to $1$\,$-$\,$5$ kpc, and, therefore, is sensitive to bright IPs at these distances. Of the 928 X-ray sources identified by the DGPS \citep{DGPS}, only 73 sources are confidently classified, 10 of which are CVs (6 being IPs).

We selected unclassified sources with an XRT brightness of $>$\,$10^{-12}$ erg cm$^{-2}$ s$^{-1}$ for our follow-up campaigns. There are 151 sources satisfying this criteria (the majority of which are classified, see Figure 9 of \citealt{DGPS}), and we observed $\sim$\,$12$ sources through our ToO programs. Three of these targets (as yet) were classified as magnetic CVs. This may suggest that there is a significant population of magnetic CVs existing in the Survey footprint, but the lack of follow-up observations (and classifications) for a more significant number of sources precludes us from drawing strong conclusions on the overall population from the classified population. The classification of our other targets is underway.

\section*{Acknowledgements}

B.~O. would like to acknowledge the significant contribution of Nicholas Gorgone to DGPS operations and analyses during his dissertation research. 

B.~O. and C.~K. acknowledge partial support under NASA Grants 80NSSC20K0389, 80NSSC19K0916, 80NSSC22K1398, and 80NSSC22K0583, and \textit{Chandra} awards GO9-20057X and GO2-23038X. 
J.~H. acknowledges support from NASA under award number 80GSFC21M0002. 

This work is based on observations obtained with the Southern African Large Telescope (SALT) under the programme 2012-2-LSP-001 (PI: D.A.H.B). Polish participation in SALT is funded by grant No. MEiN nr 2021/WK/01. 
Based on observations obtained with XMM-Newton, an ESA science mission with instruments and contributions directly funded by ESA Member States and NASA. 
This work made use of data from the NuSTAR mission, a project led by the California Institute of Technology, managed by the Jet Propulsion Laboratory, and funded by the National Aeronautics and Space Administration. 
This research has made use of the NuSTAR Data Analysis Software (NuSTARDAS) jointly developed by the ASI Space Science Data Center (SSDC, Italy) and the California Institute of Technology (Caltech, USA). 

The scientific results reported in this article are based on observations made by the Chandra X-ray Observatory.
This research has made use of software provided by the Chandra X-ray Center (CXC) in the application package CIAO. 
This work made use of data supplied by the UK \textit{Swift} Science Data Centre at the University of Leicester. This research has made use of the XRT Data Analysis Software (XRTDAS) developed under the responsibility of the ASI Science Data Center (ASDC), Italy.
This research has made use of data and/or software provided by the High Energy Astrophysics Science Archive Research Center (HEASARC), which is a service of the Astrophysics Science Division at NASA/GSFC.
This research has made use of the \texttt{VizieR} catalogue access tool, CDS, Strasbourg, France (DOI : 10.26093/cds/vizier). The original description of the \texttt{VizieR} service was published in 2000, A\&AS 143, 23 \citep{2000A&AS..143...23O}.

\section*{Data Availability}

The data underlying this article will be shared on reasonable request to the corresponding author.



\bibliographystyle{mnras}
\bibliography{magnetar-bib} 








\bsp	
\label{lastpage}
\end{document}